**Title:**

# Mobile Health Solution for College Student Mental Health: Interview Study and Design Requirement Analysis

**Authors List:** Xiaomei Wang, Alec Smith, Bruce Keller, Farzan Sasangohar

**Abstract:**

Background: Mental health problems are prevalent in college students. The COVID-19 pandemic exacerbated the problems, and created a surge in the popularity of telehealth and mobile health solutions. Despite that mobile health is a promising approach to help students with mental health needs, few studies exist in investigating key features students need in a mental health self-management tool.

Objective: The objective of our study was to identified key requirements and features for the design of a student-centered mental health self-management tool.

Methods: An interview study was first conducted to understand college students' needs and preferences on a mental health self-management tool. Functional information requirement analysis was then conducted to translate the needs into design implications.

Results: A total of 153 university students were recruited for the semi-structured interview. The participants mentioned several features including coping techniques, artificial intelligence, time management, tracking, and communication with others. Participant's preferences on usability and privacy settings were also collected. The desired functions were analyzed and turned into design-agnostic information requirements.

Conclusions: This study documents findings from interviews with university students to understand their needs and preferences for a tool to help with self-management of mental health.

## Introduction

About 1 in 3 first-year college students screen positive for at least one mental health disorder, with low variability across countries and demographics [1]. Despite 20% increased spending in university mental health services [2] and increased intention and utilization of mental health services by students, college student mental health continues to decline [3]. Poor student mental health has resulted in suicidal ideation [4], low class engagement, low GPA [5], and greater likelihood of dropping out [6]. All of these problems have been exacerbated by the ongoing COVID-19 pandemic at the time of this study, with studies showing evidence of students not effectively coping [7,8].

While students generally prefer in-person counseling [9], COVID-19 has created a surge in the popularity of telehealth services due to its ability to reach people in isolation and prevent the spread of illness [10]. In-person counseling has many disadvantages that may make alternatives like virtual counseling and mental health apps more appealing even after the pandemic. Counseling is associated with higher financial costs and is not preferred by those with self-stigma regarding mental health [9]. It is well-documented that a sizable portion of students who committed suicide never contacted their institutions' counseling center [11]. Furthermore, long wait times are the reality in many colleges with an average wait of 6 business days for a triage session and 9 business days for their first session after triage [12].

In contrast, digital interventions such as a mobile health (mHealth) application can be accessed on demand and almost instantaneously and as frequently as a user desires. Despite there being hundreds of mental health apps available on Google Play and the Apple App Store, relatively little is known about their effectiveness, as most marketing for apps is based on testimonials and unsupported claims. Less than half of these apps are supported by studies [13], and those studies have been generally criticized as biased by high rates of participant dropout [14]. Further, users tend to drop off in usage after a short period of time [15]. While there is certainly need for mHealth apps as alternative delivery of mental health care, rigorous design and evaluation methods should be taken to ensure that the apps are designed to be user-centered, evidence-based, well-integrated with professional care, and used sustainably [16].

Despite these shortcomings, mHealth presents a promising platform to support students with mental health needs, especially given the popularity and ubiquity of smartphones among college students [17]. As digital natives, students are more proficient at communicating across virtual platforms [18] which may help in managing their mental health through apps. Other research has also found that asking student patients to use a mental health app between the time they set up and attend an appointment with a counselor led to moderate app usage and better mental health outcomes [12]. In spite of these promising findings, study of mental health apps for college students  have been focused on efficacy with self-selecting participants [19], with a general gap in documented studies utilizing user-centered design methods to inquire college students' preferences, expectations, and needs to ensure prolonged use and adoption.

To address this gap, this qualitative study identified key requirements and features for the design of a student-centered, mental health self-management tool. Specifically, this paper presents emergent themes from interviews with 153 university students. The information gathered was analyzed using the Functional Information Requirement analysis method [20] to produce information requirements and design implications.

## Part I Interview Study to Understand User Needs and Preferences

A series of interviews were conducted to understand college students' needs and preferences for a mental health self-management tool.

### Methods

Semi-structured interviews were conducted with undergraduate and graduate students at a large university in South Central Texas. All interviews were conducted virtually during the Spring 2020 semester (April) due to the ongoing COVID-19 pandemic. The interview protocol was approved by the university's institutional review board (IRB) prior to data collection. Students were recruited using word of mouth (verbally or through email with a recruitment script) and were not compensated for participation. Inclusion criteria included 1) Being 18 years of age or older, 2) being a college student enrolled in the University, 3) self-identifying as having a mental health issue, 4) having the ability to communicate in English, and 5) being willing to voluntarily discuss approaches and technologies for improving student mental health.

#### Participants

A convenient sample of one hundred and fifty-three students were recruited for participation in the study; none of the participants refused to participate or dropped out of the study. After completing the informed consent process, the participants completed demographic forms.

#### Research Team

Twenty-one undergraduate students, trained in interviewing skills, conducted the interviews and had established relationships (as peers) with the participants prior to the study. One PhD student (AS), and two undergraduate students (PA, DD) conducted the analysis. All three coders were experienced in qualitative data analysis.

#### Data Collection

The interviews were all conducted virtually via telephone, Zoom [21], or Google Meet [22] during the Spring 2020 semester. The audio from the interviews was recorded and transcribed using Otter.ai [23]. The undergraduate students edited their transcripts to remove any errors. The interview protocol covered the following topics: stress, the effect of COVID-19 on mental health, general life, coping mechanisms, barriers and enablers to mental health treatment and preferences for a tool to help cope with mental health. Under each main topic, several probing questions were asked of the participants. The full interview protocol is provided in Appendix A. Due to the nature of semi-structured interviews, not every question was asked and not every question received a response.

#### Data Analysis

The transcribed interviews were analyzed using thematic analysis [24] in three phases: initial coding, focused coding and thematic coding. The three coders created initial codes independently based on the interview protocol. They then met to discuss the codes and built consensus on initial codes for the analysis. The coders then performed initial coding on 5 transcripts independently.

This phase consisted of reading through the transcript and placing responses into the corresponding codes. Following the initial coding of these 5 transcripts, they met to discuss differences and resolved discrepancies though detailed discussions. Independent focused coding of the 5 transcripts followed in which new codes were created and coded responses were moved to other codes as needed. The coders then met to discuss any differences in their focused coding and resolve discrepancies. After the focused coding discussion, the coders began coding different documents, meeting on a daily basis to update their progress, propose new codes and discuss saturation of codes. After 5 days of coding, the coders reached saturation with 153 coded transcripts. Thematic coding commenced after the focused coding process which consisted of identifying themes and ideas bridging the different codes. The coders created an initial list of themes and then discussed them with the other authors, creating a final list of themes presented in this paper. While stylistic differences may exist in the coding process, there were no major departures from the methodology employed by coders. All analysis was completed using MAXQDA 20 and MAXQDA 12 [25].

## Results

A total of 153 college student participants were interviewed. The majority of participants (n=95) were female (62%). The sample was evenly distributed among undergraduate classifications with only one graduate student: 20.26% (31) freshman, 19.61% (30) sophomores, 32.67% (50) juniors, 26.79% (41) seniors, 0.65% (1) graduate students. The students' mean age was 20.25 (SD = 1.21) years.

Table 1 presents the functions and specific features needed in a mental health management tool based on the qualitative data analysis performed on the 153 interviews.

Table 1. High-Level and Low-Level Functions and selected participant quotes

| Function | n(%) | Example Quotes |
|---|---|---|
| **Coping Techniques** | | |
| • Positivity | 67(44) | "I really think just having that daily notification with just a nice quote or something was really important for me and it's helped me a lot and kind of just start my day off just reading something that is positive and I would really love to have an app like that continues to do that." |
| • Coping Techniques/ Ways to Manage Stress | 46(30) | "I think maybe like it could give me some options with how to cope with that stress and anxiety…" |
| • Breathing Exercises | 36(24) | "Meditation and like breathing techniques, just to help calm you down and stuff like that." |
| • Calming Sounds | 34(22) | "Maybe including sounds like waterfalls. I know that some sounds are calming along with pictures to look at are calming so going to the beach." |
| • Meditation | 33(22) | "...maybe a meditation type thing where it instructs you on like breathing and ways to relax because I think a lot of people always say they want to relax but no one puts time into it so I think if there's like an app that tells you 'Do this' then maybe they'll relax some more." |

| Function | n(%) | Example Quotes |
|---|---|---|
| • Distraction Tasks | 15(10) | "...just to say you're always kind of doing something to keep your mind busy. You're not just sitting there with your thoughts." |
| **Tracking** | | |
| • Data Display | 62(41) | "Probably make a nice little line graph and track what days I was experiencing the most stress and then I could identify what was causing the stress and then try to better cope with that." |
| • Data Logging | 47(31) | "Make sure that I'm keeping up. Also, something to keep track of like reoccurring habits, stuff that's not necessarily a goal was something that you want to implement for the long term. So, whether that's meditation or exercise, or not drinking as much coffee or whatever you want it to be." |
| • Data Tracking | 44(29) | "The sensor can probably measure your sweat, measure your heart rate, and know some details we wouldn't be able to know." |
| • Journaling | 16(10) | "I'd say an easy one is a journal or a sketchbook, something tangible that I can write down or draw on, you know, what I'm feeling or what I'm dealing with, or kind of just like express emotion into and sometimes it's just kind of that expression can be a sense of a release...So that's a big thing for me." |
| • Self-assessment | 13(8) | "...some quizzes where you can go through and figure out what's wrong with you...maybe that app has that kind of feature or something and then at the end, once it does give you a result, also gives you like tips to help with that specific thing." |
| **Smart Features** | | |
| • Coaching | 32(21) | "Tell me if I'm anxious and tell me whether or not things I should do to like help with my anxiety." |
| • Virtual Assistant | 26(17) | "...maybe even like have an AI implemented to learn about your responses and learn about what stresses you out and give mindful suggestions about how you can improve." |
| • Prediction | 5(3) | "I don't know if this is possible, but like maybe predict future stress and anxiety based on like previous pattern[s]." |
| **Time Management** | | |
| • Planners | 28(18) | "I guess like a planner, time organizer, that organizes things and reminds you." |
| • Context-aware Reminders | 24(16) | "...feature that says like, you've been scrolling for far too long. Go get some water, take a walk, and then come back later." |
| • Calendar | 16(10) | "I would say, probably, like a calendar just to manage my time." |
| **Communication** | | |
| • Communication | 18(12) | "It would be really great to be able to just either live chat with somebody or some just some kind of communication with an actual person within it." |
| • Counselor Services | 10(7) | "I think it would be helpful to have access to like a licensed counselor within an app." |

## Coping Techniques

Participants provided a variety of coping techniques that they would like to see in the tool.

***Positivity.*** The participants believed that an effective self-management tool should enforce positivity. Some participants (13/153; 8.5%) wanted the tool to contain content that results in "happiness". For instance, some participants wanted pictures of cute animals, reminders of people or pets they love or stories that made them happy. In addition to things that make them happy, some participants (25/153; 16.3%) wanted the tool to provide positive affirmation. Some of the suggestions for this feature included encouragement to persevere, motivational quotes and motivational stories.

***Coping Techniques/ Ways to Manage Stress.*** Participants (46/153; 30.1%) stated their preference for coping techniques that would help them calm down, relax or manage stress. Rather than providing specifics, the participants generally stated their desire for coping techniques or other proven methods of coping to be present on the tool.

***Breathing Exercises.*** Some participants (36/153; 23.5%) stated that having validated breathing exercises on the tool would help them with their stress. They also thought that the breathing exercises would help them calm down or focus.

***Calming Sounds.*** Participants (34/153; 22.2%) wanted an audio feature that would help them relax. Within the broad category of calming sounds, some of the participants (21/153; 13.7%) wanted the tool to have on demand access to calming music. While some participants stated explicitly that they would want the capability to add and listen to music they like, others just wanted generic, relaxing music. Aside from relaxing music, some of the participants (13/153; 8.5%) thought that calming sounds such as rain, white noise or sounds from nature would be a good addition to the tool. A couple of the participants (3/13; 23.1%) also thought that spoken word that evokes the autonomous sensory meridian response (ASMR) would also be a good method for calming down.

***Meditation.*** Some of the participants (33/153; 21.6%) mentioned that the presence of a meditation feature on the tool would be helpful for calming them down, focusing or increasing mindfulness; a few particular additions to the meditation feature included voice-guided meditations and Yoga.

***Distraction Tasks.*** Some participants thought that the tool would be useful for a momentary respite from stress in the format of games and other distractions. Several participants (6/153; 3.9%) mentioned that playing games on the tool could help take their mind away from stress in their life. Specifically, some participants wanted the ability to play their favorite games. Several others (9/153; 5.9%) thought that the tool would be helpful if it could provide general distractions from stressful events in their life.

## Tracking

Participants wanted their information, such as heart rate, steps, moods, stress moments, etc., to be tracked and logged on to the tool to improve self-awareness, track triggers, show health trends, and enable predictions.

***Data Display.*** Participants (62/153; 40.5%) were interested in displays presenting their current mental health state including physiological variables such as heart rate and breathing rate, as well as stress levels. In addition, participants expected a feature that displays the collected data in the form of charts and trendlines. Some participants showed their preference for options to view their trends arranged by day, week, and month.

***Data Logging.*** Participants (47/153; 30.7%) preferred some form of a tracking feature built into the tool. The users expressed interest in manually logging information such as sleep, eating habits, medication, stress moments, amount of water consumed, exercise, etc. to maintain their lifestyle.

***Data Tracking.*** Participants (44/153; 28.8%) wanted the tool to automatically track and record high stress or trigger points. They thought that this could be accomplished by measuring and analyzing heart rate and breathing data or other physiological data. They also wanted the tool to report these stress moments in a compassionate way.

***Journaling.*** Some participants (16/153; 10.5%) stated that journaling would be a helpful feature as they were habituated to journaling whenever they are stressed. They preferred to have a space within the tool for journaling or freely expressing their thoughts or feelings about a stressor or about any other aspect they like to journal about or log their feelings/emotions within the tool.

***Self-assessment.*** Some participants (13/153; 8.5%) also stated that they wanted to fill out a self-assessment questionnaire periodically (typically once a day). The questionnaire would have questions about the user’s day, stress moments, current mood, and their stress levels. Participants expected to have access to information about their progress over time and information about how to improve.

## Smart Features

Participants mentioned wanting the tool to have advanced computational, conversational and smart features to help them cope with stress.

***Coaching.*** Participants (32/153; 20.9%) stated their preference for the tool to be able to tell them what to do when they are stressed with limited prompting. The scenarios in which some of the participants envisioned this feature to function include detection of stress and emotions or other feelings that may indicate stress and populating events or activities on their calendar. Participants mentioned that tool should direct them through necessary tasks to mitigate mental health issues. Several participants (22/153; 14.4%) wanted the tool to recommend immediate actions to reduce their heart rate and stress levels.

***Virtual Assistant.*** Participants (26/153; 17.0%) wanted a virtual assistant. Some of these features included chatbot features (2/26; 7.7%), and advanced notifications and planning that will help participants remember events (4/26; 15.4%).

***Prediction.*** Some participants (5/153; 3.3%) mentioned predicting the occurrence of stress using a variety of measures: heart rate or other physiological markers, events on the calendar, journal entries, emotions and any other data that is provided to the tool by the user. They hoped that the prediction of stress would reduce the actual occurrence of stress because they would be able to mitigate it with coping techniques available on the tool.

### Time Management

Participants acknowledged that poor time management is a major stressor and favored features that can help them manage their time in the form of reminders, scheduling, and planning their day, week or month.

***Planners.*** Some participants (28/153; 18.3%) wanted the tools necessary to effectively manage time and their stress. They wanted a holistic time manager or a scheduler for themselves that guides them through organizing their schedules. They thought that the tool would even be able to build the schedule for them.

***Context-aware Reminders.*** Some participants (24/153; 15.7%) stated that they felt stressed out whenever they were overwhelmed with tasks and could not keep up with them and wanted a reminder to help them. Several participants wanted an option within the tool to set reminders for their daily tasks to stay on track with their lives. Some of the specific cases they brought up included phone usage, taking breaks, walking, finishing assignments or staying motivated.

***Calendar.*** Several participants (16/153; 10.5%) said that they wanted a calendar feature to plan their day or week. Since college participants are involved in more than just taking classes, it was important for them to plan their time efficiently to perform well in all areas and avoid stress.

### Communication

Participants mentioned wanting to be able to communicate thoughts and information with others through the tool.

***Communication.*** Some of the participants (18/153; 11.8%) wanted to be able to communicate with other users through the tool. This largely consisted of wanting to talk to other users of the tool via live chat because they would be able to understand what the other users were going through.

***Counselor Services.*** In addition to communicating with other users, some participants (10/153; 6.5%) mentioned wanting to communicate with a counselor through the tool. This was brought up either as having someone available to speak to anonymously or being able to reach a counselor that the user has a relationship with to talk about things as needed.

Table 2 presents the desired characteristics of a mental health mobile application based on the qualitative data analysis on the interviews. Two major themes of characteristics emerged from the interviews: usability and privacy/data sharing.

Table 2. Mental Health App Characteristics and Selected Participant Quotes

| Expected Characteristics | n(%) | Example Quotes |
|---|---|---|
| **Usability** | | |

| Expected Characteristics | n(%) | Example Quotes |
|---|---|---|
| • Reliability | 19(12) | "I [have] worn a heart monitor before and it's hard for doctors or whoever to differentiate between, you know, when you're watching a scary movie or when you're stressed or when you're exercising and all that." |
| • Calming | 17(11) | "Color wise obviously nothing too intense. I don't know what those colors would be but not too much contrast. If calm could be a color." |
| • Wearable | 12(8) | "For me to most likely be a wearable sensor like an apple watch or something like that, along those lines." |
| • Customizable | 10(7) | "I think it would be cool if you could [...] personalize the vibe of the app, like the colors that you want. Like if the user picked out the colors at the start right when they signed up for the app or downloaded the app…" |
| • Easy to Use | 9(6) | "If we're talking technology, then I would have some sort of app that's very, very simple." |
| • Promote Sustainable Use | 4(3) | "...like encouraged me to use the app." |
| • Non-time Consuming | 2(1) | "That wouldn't take me too much time to be able to go into maybe." |
| **Privacy and Data Sharing** | | |
| • Providers' Access to Data | 15(18) | "I think that would be helpful. But I think it would be like...it shouldn't do that automatically, unless you like give it permission, which, obviously. But yeah, would be helpful, I guess. I don't know if I would utilize that but I think that would be a good feature for it to have." |
| • Researchers' Access to Data | 7(8) | "I would probably expect the data to be compiled and to be sorted against other users so that the company itself can better make trends and make better suggestions to individual users about how to manage stress." |
| • Security/Privacy/ Credibility Concerns | 5(3) | "...if it's collecting my heart rate, what else is in collecting, you know…" |

## Usability

Some of the participants thought that the tool should take the form of a smartphone or smartwatch application and provided suggestions for how to make it more appealing and usable.

***Reliability.*** Some participants (19/153; 12.4%) wanted the tool to be reliable in differentiating between stress and exercise or other activities. They thought that this was important because they did not want to be engaging in physical activity and the tool accidentally recognized stress due to elevated heart rate; they thought that the mischaracterization of stress would in turn increase their stress upon later review of their stress levels.

***Calming.*** Participants (17/153; 11.1%) suggested that the app use calming colors and avoiding colors that are too vivid in its design to make people more at ease using it. Several participants who suggested calming colors be used (6/17; 35.3%) suggested green or blue specifically as they considered those more calming.

***Wearable.*** Some of the participants (12/153; 7.8%) brought up the idea of using the app on wearable technologies and in particular, smartwatches. Some of them mentioned this as part of making it easily accessible to use for people that already own smart watches. Two participants mentioned that wearable technology for this use could be made to be innocuous and make it less likely others could identify the purpose of the technology was to treat mental health, which they saw as important.

***Customizable.*** Another suggestion some participants (10/153; 6.5%) had was to make the app customizable. They wanted to be able to change, among other things, the colors, the layout, and the features that were used to make it feel more personalized.

***Easy to Use.*** Some of the participants (9/153; 5.9%) mentioned that the tool should be easy to use. They wanted it to be accessible to many people by not having a complicated user interface. They also thought that if it was easy to use, it would induce less stress to the users.

***Promote Sustainable Use.*** A few participants (4/153; 2.6%) mentioned that they would like the app to encourage them to use it.

***Non-time Consuming.*** A couple of participants (2/153; 1.3%) mentioned that they would not want the app to be time consuming. They did not want it to take long for them to use it or they would not use it as often.

## Privacy and Data Sharing

Some participants mentioned their preferences on privacy and data sharing. 85 participants were specifically asked their preference on sharing data to providers or researchers.

***Providers' Access to Data.*** In response to being asked what they would expect the tool to do with their mental health data, several participants (13/85; 15.3%) mentioned they would want the option to send it to a healthcare provider. This response was given for several reasons, including wanting their data analyzed for any alarming trends or patterns, wanting their doctor to have the data because they thought it would be helpful, and wanting it to seek medical help for them if the values are deemed clinically significant. Some users mentioned such transmission of data should not be automatic and the user should opt in to sending the data. In contrast, there were two participants (2/85; 2.4%) who said they would not want healthcare provider have access to the data because of privacy or reliability concerns.

***Researchers' Access to Data.*** In addition to being asked about their preferences to send data to healthcare providers, participants were also asked about sending the same data to researchers and several (7/85; 8.2%) responded that they would as long as the data was sent anonymously and their identity was protected.

***Security/Privacy/Credibility Concerns.*** Several participants (5/153; 3.3%) brought up privacy as something they were concerned about. These participants were concerned about the app collecting “so much information” on them and wanted assurances that the data would not be given to third parties or used improperly.

# Part II Functional Information Requirement Analysis

Functional Information Requirement (FIR) analysis (Khanade et al., 2018) was conducted utilizing the desired functions, identified from the interview (Table 1), to derive information requirements (IRs) for a mental health self-management app. IRs are solution-neutral, design-independent requirements that can be used as an objective assessment of whether user needs were met for design.

## Methods

### Research Team

Nine analysts conducted the FIR analysis, including two postdoctoral researchers (SH, XW), one PhD student (AS), one Master’s student (BK), and five undergraduate students (ZX, ER, FI, AC, DD). Team members had expertise in one or more areas of the following: qualitative analysis, usability testing, interface design and mobile app development.

### Data Analysis

The whole team of analysts had an initial discussion to train themselves on the FIR analysis methods. Results from Part I, specifically Table 1 (desired functions) was used as initial input for the FIR analysis. Lead analyst of the interviews, AS, introduced the themes (high-level function) and subthemes (low-level function) to the whole team. Team members asked for clarifications until each member fully understood the user needs collected from the interview. The whole team then analyzed a single function together to form consensus and to standardize the process.

Team members were then randomly assigned to five groups, each containing three to five analysts, to analyze the five high-level functions. In each group, each analyst first identified some IRs independently. Then the group discussed to merge the IRs into a combined list.

After all groups finished their analysis, the whole team came together to review all IRs. Each group presented the results. Members from other groups asked for clarification, and suggested edits. The team also evaluated the feasibility and priority for the development purposes. The feasibility was mainly evaluated by the team member with expertise in mobile app development. The evaluation on priority was based on both the amount of evidence (e.g., frequency of functions mentioned in the interviews), and the feasibility assessment.

## Results

Table 3 shows the information requirements identified for all the functional needs. The left-hand column shows the function, or what is to be performed (by the tool or by the user) and the right-hand column shows the information requirements, or what information is needed (by the tool or user) for the function to be performed. All of the functions can be mapped back to the themes and subthemes identified from the thematic analysis described in Part I. These functions and information requirements can be used for the design and development of interactive interfaces.

Table 3. Functional Information Requirements for a College Student Mental Health Self-Management Mobile App

| Function | Information Requirements |
|---|---|
| **Coping Techniques** | |
| • Positivity | 1) Choice of quotations/sources (e.g. Philosophers/Humor/Scriptural)<br>2) Bank of quotes<br>3) User's preference on notification time/form<br>4) Messages that the user has saved that were written by the user<br>5) Task accomplishments |
| • Coping Techniques/Ways to Manage Stress | 1) Notifications of stress levels;<br>2) Decide appropriate intervention<br>3) Suggestion of relaxation activity/technique (Environment noisy/quiet; Time Available for Relaxation; Whether User Allow Video/Audio)<br>4) Feedback<br>5) Set break times<br>6) Activity tracking and duration |
| • Breathing Exercises | 1) Guidance of steps - voice, video, pictures, text<br>2) Choice of length<br>3) Breathing time for inhalation and exhalation<br>4) Visual representation of inhalation and exhalation<br>5) Tutorial, ambient sound<br>6) Number of breaths set by user |
| • Calming Sounds | 1) Collection of audio bits |
| • Meditation | 1) Guidance of steps/tutorial - voice, video, pictures, text<br>2) Choice of length<br>3) Choice of music - themes, instrumental/vocal, silent<br>4) Summary report: Average heart Rate, duration<br>5) Self-assessment: quality of meditation/relaxation |
| • Distraction Tasks - Games | 1) List of games;<br>2) Game type (e.g. hand-eye coordination/focus/puzzle)<br>3) Easy games that do not take a lot of thinking;<br>a) Showing a grid of images and then matching each image;<br>b) Simon says;<br>c) Rhythmic tapping (similar to the exercise in First Watch Device)<br>4) Difficulty level;<br>5) Previous scores;<br>6) Previous progress (for paused games) |
| • Distraction Tasks - Other | 1) Identify stressor type<br>2) Tutorials on distraction (e.g. "look around the room and describe in detail what you see for 3 minutes")<br>3) A list of leisure activities that the user likes to do |

| Function | Information Requirements |
|---|---|
| **Tracking** | |
| • Data Display | 1) User's HR, stress, BP<br>2) Healthy Range<br>3) Previous Days Information<br>4) Display of Information<br>5) Time Measured<br>6) Data Analysis and Interpretation (based on demographics) |
| • Data Tracking - Physiological Vitals | 1) User HR/Blood Pressure<br>2) Measurement Time<br>3) Previous Days Data<br>4) Display of Information<br>5) Scale of HR<br>6) Device Availability |
| • Data Logging - Emotions | 1) Date<br>2) User's emotions/mood<br>3) previous days inputs |
| • Data Logging - Sleep | 1) Date<br>2) Time Fell Asleep/Woke Up (Total Time Slept)<br>3) User's assessment of amount and quality<br>4) previous days input |
| • Data Logging - Food Intake | 1) Date<br>2) Time<br>3) User's Food Eaten (Type, Amount, user's assessment of whether amount/time was appropriate)<br>4) Previous Days Information<br>5) Display of Information |
| • Data Logging - Trigger/High Stress | 1) Date<br>2) User's self-assessment (scale score, moods)<br>3) Objective assessment (e.g. HR-based)<br>4) Current time<br>5) Stress level measured or perceived (e.g. PSS)<br>6) Source of stress (e.g. work, academics, family)<br>7) Previous days inputs |
| • Journaling | 1) Current Date<br>2) User's Notes<br>3) Previous Days Inputs |
| • Self-assessment | 1) Scales<br>2) Previous Days Scores<br>3) Display of Information<br>4) User's Preference on Evaluation Frequency |
| **Smart Features** | |

| Function | Information Requirements |
|---|---|
| • Coaching | 1) User's stress factors (HR, physiological, emotions, notes, etc.)<br>2) AI commands |
| • Virtual Assistant - Chatbot | 1) User input (comments, feelings, emotions, etc.)<br>2) AI responses |
| • Virtual Assistant – Advanced Notification and Planning | 1) User inputs from various other features (calendar, reminders)<br>2) AI information concerning upcoming event |
| • Prediction | 1) User's stress factors (HR, physiological, emotions, notes, etc.)<br>2) Information of impending stress event |
| **Time Management** | |
| • Planners | 1) User's Tasks<br>2) Relationship between Tasks<br>3) Groups/Categories; Labels |
| • Context-aware Reminders | 1) Current Date<br>2) Planned Date<br>3) User Activity<br>4) Time Zone<br>5) Time of Reminder<br>6) Modality of Reminder<br>7) Repetition |
| • Calendar | 1) Current Date<br>2) Planned Date<br>3) User Activity<br>4) Time Requirement/Duration for Activity<br>5) Time Zone |
| **Communication** | |
| • Communication | 1) User's input (comments, posts, pictures, etc.)<br>2) Other user's input<br>3) Usernames<br>4) Platform to communicate/Private messaging |
| • Counselor services | 1) Care Provider's Contact Info<br>2) Care provider preferred method for contact<br>3) User's contact info<br>4) User's preferred method for contact<br>5) Message for care provider<br>6) Message for user<br>7) Physical Office Address<br>8) Locations on a map<br>9) Specialization/skills<br>10) Patient reviews/ratings<br>11) Insurances accepted |

Figure 1 shows one example of the displays designed for a mobile health application based on the FIR analysis. A low-level function, Data Display (details in Table 1) was analyzed by three team members. Each member individually analyzed and listed the IRs for the function. During group discussion, the three lists were combined as a final list (as is in Table 2). The two screenshots show how the IRs were considered during design. On the left, a daily summary was shown with user's data (e.g., 114 BPM average heart rate, IR 1) and time measured ("May 19, 2020", IR 5). Clicking on the "Heart Rate" leads to the screenshot on the right that shows detailed information – information that are on the list but not shown in the summary. For example, the chart shows data of the past week (IR3) and serves as a special display of information (IR4); the weekly average involves data analysis (IR6); and the tables about heathy rest heart rate ranges at the bottom help users interpret the data (IR6). The IRs list guides the designer to check whether all users' needs were met on this function.

**High-level Function:** Tracking
**Low-level Function:** Data Display

**Information Requirements Identified by 3 Individual Team Members:**

| Member 1 | Member 2 | Member 3 |
|---|---|---|
| • Data Analysis and Interpretation; | • User Heart Rate; | • User's HR, stress, etc.; |
| • Healthy Range; | • User Blood Pressure; | • Previous days levels; |
| • Demographics | • Clock | • Display of information |

**Combined List of Information Requirements:**

1. User's HR, stress, etc.;
2. Healthy Range;
3. Previous Days Information;
4. Display of Information;
5. Time Measured;
6. Data Analysis and Interpretation (based on demographics)

**Example Displays Designed Based on the Above Analysis**

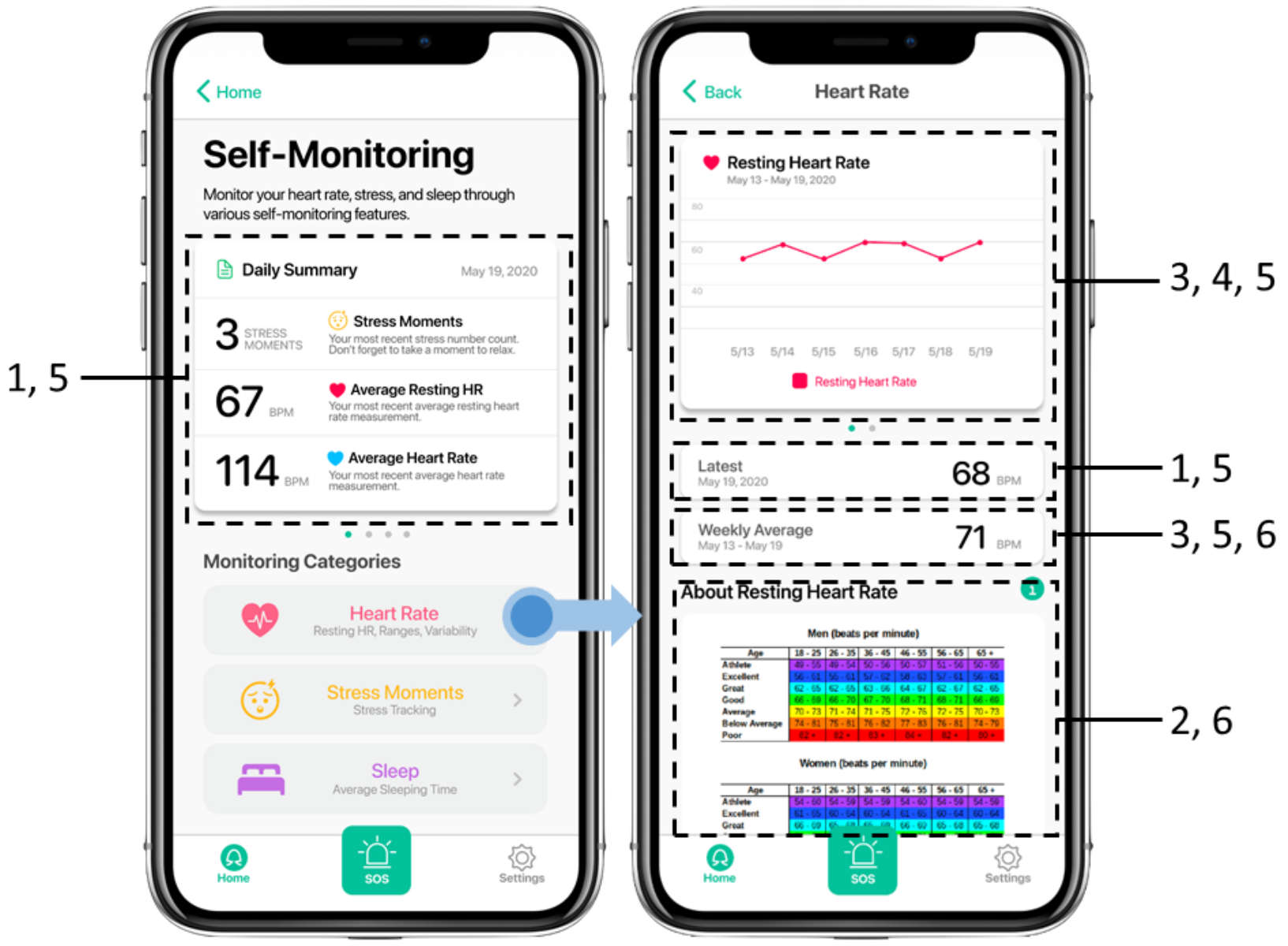

Figure 1: FIR Analysis and Display Design for a Low-Level Function, Data Display.

# Discussion

This study sought to elicit design requirements for a mental health self-management technology from college students. The themes identified from the analysis demonstrate the broad spectrum of requirements and expectations students have for a mental health technology including coping techniques, artificial intelligence features, tracking features, time management help, communication, and good design. These themes were then used to determine the functions and information requirements of a mental health technology using the FIR framework.

## User Requirements for a College Student Mental Health Self-Management Tool

***Coping techniques.*** Our findings suggest that most common feature of a mental health mobile application is to provide effective coping techniques or tools. Many of the preferred coping methods documented in this paper are in line with the previous work showing their efficacy in reducing stress and anxiety including: meditation [26,27], breathing exercises [28,29], calming sounds [30,31], distractions [32], video games [33], and positivity [34,35]. Some of these coping mechanisms have also been evaluated or implemented across different digital technologies, such as meditation [36], reading [37], breathing exercises [38], calming sounds [39], and video games [40]. However, despite the promise shown by these methods to improve self-management of mental health, to our knowledge, there is a general gap in tools that provide all or majority of these features [16,41]. Given varying user preferences, providing on-demand access to wider range of coping tools or methods may serve as an effective and inclusive approach to enabling mental health self-management among students.

***Tracking.*** A major theme among participants was a preference for the inclusion of a mental health tracking feature in the tool. Among features expected by students, tracking of daily activities has shown to be an effective method to reduce stress and improve mental health [42,43]. In particular journaling has shown to reduce levels of stress among university students [44]. The participants also thought that having on-demand access to information about their current mental health state would be beneficial to maintain awareness of their condition and motivate self-management. Previous research has shown that the availability of personal health information in one place allows for users to adapt their behavior to healthier habits and easily recognize changes while also aiming for goals measured and collected by various wearable technologies [45]. Currently-available technologies can capture many types of information that participants would want tracked including heart rate [46]; emotions [47]; activity tracking [48]; and notetaking to track daily activities including sleep, food, and any other daily tasks [49].

***Smart Features.*** The findings of this research suggests that students expect mental health tools to provide smart features that act as a companion and help them in carrying out their tasks to minimize stress levels. One feature desired by a few participants was a chatbot. While relatively novel, chatbots have shown effective in helping people manage or improve their mental health [50,51]. In addition to chatbots, assistive technologies, in the form of context-based commands and assistance with everyday tasks, also show promise for those with mental disabilities and cognitive impairment [52,53]. Participants also wanted the tool to use their physiological characteristics to help reduce stress or predict stress events while distinguishing between true stress events and other activities. This is in line with recent work [54,55] evaluating machine learning algorithms that used heart rate and accelerometer data for the recognition of stress events among combat veterans. Other studies have also shown the ability to distinguish between activities using kinematic data, suggesting the feasibility of identifying stress moments more accurately [56,57]. With the advancement of technology that may be able to accommodate the desired functions of participants, artificial intelligence may be a feasible addition to a mental health tool for stress reduction.

***Time management.*** Many of the participants wanted the tool to help them with time management. Of the features that they mentioned, general time planning was the most desired followed by features like calendar and reminders. Time management activities have been shown to reduce stress among workers and students [58,59]. A calendar feature can help students better organize their activities in a more visible manner and prioritize tasks by using effective time management principles. Additionally, participants also wanted a feature with reminders. The reminders would work in conjunction with the calendar to actively remind users of important tasks or events, relieving the user of the burden of constantly having to keep a mental track of tasks and schedules. The aforementioned features, calendars and reminders, are already present in digital technologies making them easily accessible for most people [60]. To better support mental health self-management needs, these features could be integrated with other desired features. For instance, self-mental health assessments could be integrated with the calendar to automate scheduling and reminders to complete assessments or an AI can analyze calendar information and health data to identify key stressors and make suggestions for optimizing time management.

***Communication.*** Some participants mentioned their desire for the tool to have a communication feature. Regardless of the social stigma associated with mental health, participants still wanted the ability to communicate with others about their stress and anxiety [61]. Some studies have shown that talking out about one’s own mental health with people outside of professional therapy, like family and friends, reduces stress and anxiety [62]. Still, talk therapy, or speaking to a licensed therapist, is helpful in reducing stress and anxiety and may even be effective over digital technologies [63]. Overall, the ability to communicate with others, both professionals and friends/family members, may help reduce stress and anxiety and improve mental health.

***Usability.*** Some participants emphasized their need for the tool to be usable. Similar requirements on self-management tools have emerged from many studies [64]. For example, a diabetes self-management study revealed user needs for the tool to be reliable and easy to use, which were also identified in the current study [65]. Usability has also been found to correlate with user rating in app store [16]. However, few mental health tools developed were evaluated through documented usability testing [66], which often leads to poor adoption and engagement [67].

***Privacy and Data Sharing***. Several participants mentioned privacy concerns especially when physiological data is collected automatically by the tool. Participants also showed divergent preferences on sharing data with health providers or researchers. Our findings suggest that data should be held secure and sharing features should provide clear options. Recent reviews found that only around half of the mental health apps on the market had privacy policy [68,69].

## Limitations

The current study has several limitations, in both the first part of data collection and thematic analysis and the FIR. First, the data was collected from a single university and the results may not generalize to all college students. More work is needed to verify the college students' needs and expectations from various colleges and universities in different geographic settings and varying size to verify these findings. In addition, the interviewed students were not assessed clinically for any mental health problems. Lastly, the qualitative analysis is subject to bias despite best efforts to control it.

## Conclusion

This study documents findings from interviews with university students to understand their needs and preferences for a tool to help with mental health. The participants mentioned several features including coping techniques, artificial intelligence, time management, tracking, and communication with others. These features were then evaluated using FIR to determine the functional and information requirements to actualize them in a tool. Future work involves applying the findings of the FIR to the design of a tool, in this case a mobile health application. Following the design, the tool will be evaluated for usability, preferences and adjusted further before eventual release on a college campus where it will be evaluated in the naturalistic setting, seeing how well it is adopted by the student population for managing stress and mental health problems.